\documentclass[12 pt, amsfonts, amssymb,color]{article}

\usepackage{amsmath,amssymb,amsfonts,latexsym,float,graphics,epsfig}

\def\R{\mathbb{R}}

\begin{document}

\newtheorem{axiom}{Definition}[section]
\newtheorem{theorem}{Theorem}[section]
\newtheorem{axiom2}{Example}[section]
\newtheorem{lem}{Lemma}[section]
\newtheorem{prop}{Proposition}[section]
\newtheorem{cor}{Corollary}[section]
\newcommand{\be}{\begin{equation}}
\newcommand{\ee}{\end{equation}}

\newcommand{\equal}{\!\!\!&=&\!\!\!}
\newcommand{\rd}{\partial}
\newcommand{\g}{\hat {\cal G}}
\newcommand{\bo}{\bigodot}
\newcommand{\res}{\mathop{\mbox{\rm res}}}
\newcommand{\diag}{\mathop{\mbox{\rm diag}}}
\newcommand{\Tr}{\mathop{\mbox{\rm Tr}}}
\newcommand{\const}{\mbox{\rm const.}\;}
\newcommand{\cA}{{\cal A}}
\newcommand{\bA}{{\bf A}}
\newcommand{\Abar}{{\bar{A}}}
\newcommand{\cAbar}{{\bar{\cA}}}
\newcommand{\bAbar}{{\bar{\bA}}}
\newcommand{\cB}{{\cal B}}
\newcommand{\bB}{{\bf B}}
\newcommand{\Bbar}{{\bar{B}}}
\newcommand{\cBbar}{{\bar{\cB}}}
\newcommand{\bBbar}{{\bar{\bB}}}
\newcommand{\bC}{{\bf C}}
\newcommand{\cbar}{{\bar{c}}}
\newcommand{\Cbar}{{\bar{C}}}
\newcommand{\Hbar}{{\bar{H}}}
\newcommand{\cL}{{\cal L}}
\newcommand{\bL}{{\bf L}}
\newcommand{\Lbar}{{\bar{L}}}
\newcommand{\cLbar}{{\bar{\cL}}}
\newcommand{\bLbar}{{\bar{\bL}}}
\newcommand{\cM}{{\cal M}}
\newcommand{\bM}{{\bf M}}
\newcommand{\Mbar}{{\bar{M}}}
\newcommand{\cMbar}{{\bar{\cM}}}
\newcommand{\bMbar}{{\bar{\bM}}}
\newcommand{\cP}{{\cal P}}
\newcommand{\cQ}{{\cal Q}}
\newcommand{\bU}{{\bf U}}
\newcommand{\bR}{{\bf R}}
\newcommand{\cW}{{\cal W}}
\newcommand{\bW}{{\bf W}}
\newcommand{\bZ}{{\bf Z}}
\newcommand{\Wbar}{{\bar{W}}}
\newcommand{\Xbar}{{\bar{X}}}
\newcommand{\cWbar}{{\bar{\cW}}}
\newcommand{\bWbar}{{\bar{\bW}}}
\newcommand{\abar}{{\bar{a}}}
\newcommand{\nbar}{{\bar{n}}}
\newcommand{\pbar}{{\bar{p}}}
\newcommand{\tbar}{{\bar{t}}}
\newcommand{\ubar}{{\bar{u}}}
\newcommand{\utilde}{\tilde{u}}
\newcommand{\vbar}{{\bar{v}}}
\newcommand{\wbar}{{\bar{w}}}
\newcommand{\phibar}{{\bar{\phi}}}
\newcommand{\Psibar}{{\bar{\Psi}}}
\newcommand{\bLambda}{{\bf \Lambda}}
\newcommand{\bDelta}{{\bf \Delta}}
\newcommand{\p}{\partial}
\newcommand{\om}{{\Omega \cal G}}
\newcommand{\ID}{{\mathbb{D}}}
\newcommand{\pr}{{\prime}}
\newcommand{\prr}{{\prime\prime}}
\newcommand{\prrr}{{\prime\prime\prime}}
\title{  An analytic technique for the solutions of nonlinear oscillators with damping using the Abel Equation}
\author{A Ghose-Choudhury\footnote{E-mail aghosechoudhury@gmail.com}\\
Department of Physics, Surendranath  College,\\ 24/2 Mahatma
Gandhi Road, Calcutta 700009, India\\
\and
Partha Guha\footnote{E-mail: partha@bose.res.in}\\
SN Bose National Centre for Basic Sciences \\
JD Block, Sector III, Salt Lake \\ Kolkata 700098,  India \\
}

\date{ }

 \maketitle

\smallskip


\smallskip

\smallskip

\begin{abstract}
Using the Chiellini condition for integrability we derive explicit solutions for a generalized system of Riccati equations $\ddot{x}+\alpha x^{2n+1}\dot{x}+x^{4n+3}=0$ by reduction to the first-order Abel equation assuming the parameter $\alpha\ge 2\sqrt{2(n+1)}$. The technique, which was proposed by Harko \textit{et al},  involves use of an auxiliary system of first-order differential equations sharing a common solution with the  Abel equation. In the process analytical proofs of some of the conjectures made earlier  on the basis of numerical investigations in \cite{SJKB} is provided.
\end{abstract}

\smallskip

\paragraph{Mathematics Classification (2010)}:34C14, 34C20.

\smallskip


\section{Introduction}
Second-order ordinary differential equations (ODEs) with linear
damping are the most commonly studied extensions of undamped
motion, the simplest example being the case of damped oscillations
$\ddot{x}+\gamma \dot{x}+\omega^2 x=0$ which admits a closed-form
solution. In the case of nonlinear ODEs even with linear damping
the construction of a closed form solution is often a nontrivial
task and such equations often display a variety of  interesting
phenomena such as chaos in the case of
non-autonomous nonlinear terms, complex periodicity, limit cycles etc. An
equation of the form $\dot{x}+f(x)\dot{x}+g(x)=0$ where $f(x)$ and
$g(x)$ are arbitrary $\mathcal{C}^\infty(I)$ real-valued functions
of $x$ defined on a real interval $I\subseteq \R$ is known as a
Li\'{e}nard equation \cite{Lienard}. There exists a vast
literature on this equation alone as it is the favored equation
for modelling  several phenomena ranging from electrical circuits,
heart beat activity, neuron activity, chemical kinetics to
turbulence in fluid dynamics \cite{VDP1,VdPol, FN,Strogatz}.
Mathematical techniques such as those of Lie symmetries
\cite{Garcia, Car1} and Wierstrass integrability have been used
to analyse the Li\'{e}nard equation \cite{GG}. Its generalization
the Levinson-Smith equation $\ddot{x}+f(x, \dot{x})\dot{x}+g(x)=0$
\cite{LS} has found applications in astrophysics where for
instance the time dependence of perturbations of the stationary
solutions of spherically symmetric accretion processes is modelled
by an equation of this form \cite{Ran}.\\

From a practical point of view the assumption of linear damping is
often insufficient. Generally it is possible to divide oscillator
systems into two broad categories, those with linear damping and
nonlinear elasticity and those with nonlinear damping and linear
elasticity. Extensive studies of both these categories may be
found in \cite{Mickens, WvH, Pilipchuk, NM}. Various quantitative
methods have been employed for their analysis depending on the
context and convenience such as the method of multiple scales,
successive approximation, averaging method besides qualitative
studies \cite{BM, Magnus, Andro}. \\

The following generalization of the Li\'{e}nard equation involving
a quadratic dependance on the velocity besides the usual linear damping term,
\textit{viz} \be\label{a}
\ddot{x}+g_2(x)\dot{x}^2+g_1(x)\dot{x}+g_0(x)=0,\ee was studied by
Bandic \cite{Bandic}. Special cases of this equation  corresponding to $g_1(x)=0$ naturally occur for oscillators
involving a variable mass and are derivable from a Lagrangian of the form
$L(x,\dot{x})= \frac{1}{2}m(x)\dot{x}^2+V(x)$.   Recently Kovacic and Rand \cite{KR}  studied
several examples of a position-dependent coefficient of the kinetic energy, which stem from a position-dependent mass or are the consequence of geometric/kinematic constraints. Some notable examples of position-dependent mass systems
 include the Mathews-Lakshmanan oscillator equation \cite{ML}, which has also been studied in the quantum
regime, the quadratic Loud systems \cite{Loud} and the Cherkas system
\cite{CG}.  In \cite{Cveticanin1} Cveti\'{c}anin analysed the case of strong quadratic
damping with a model equation given by $\ddot{x}+x+2\delta
\dot{x}|\dot{x}|=0$. The issue of isochronicity
in equations of the Li\'{e}nard type has also been extensively studied \cite{Sabatini}.\\

In \cite{SJKB} a variant of the generalized Riccati system of equations, viz
\be\label{R1}\ddot{x}+\alpha x^{2n+1} \dot{x}+x^{4n+3} =0,\ee  was considered. It was established on the basis of numerical studies that for $\alpha$ much
smaller than a critical value the dynamics is periodic, the origin being a centre.
Furthermore the solution changes from being periodic to aperiodic at a
critical value $\alpha_c=2\sqrt{2(n + 1)}$, which is independent of the initial conditions.
This behaviour is explained by finding a scaling argument via which the
phase trajectories corresponding to different initial conditions collapse onto a
single universal orbit. Numerical evidence for the transition is shown. Further,
using a perturbative renormalization group argument, it is conjectured that
the oscillator, \be\label{R2}\ddot{x} + (2n + 3)x^{2n+1}\dot{x} + x^{4n+3} + w_0^2x = 0,\ee exhibits isochronous
oscillations. The correctness of the conjecture is established numerically.
 In this communication we provide analytical proofs for some of these assertions.

Equation (\ref{a}) may be reduced to a first-order ODE by means of
the transformation $\dot{x}=1/v$, namely \be\label{b}
\frac{dv}{dx}=g_0(x)v^3+g_1(x)v^2+g_2(x)v=F(x, v).\ee This is an
Abel equation of the first-order and first kind and may be viewed
as a generalization of the Riccati equation. Such  equations, which
first appeared in  course of Abel's investigations of the
theory of elliptic functions,  usually arise in
problems involving the reduction of order of second and higher-order equations
and are frequently encountered in  modelling of practical
problems, e.g., the Emden equation, the van der Pol equation etc.
They are also relevant in the study of quadratic systems in the
plane \cite{AGG}, the centre-focus problem \cite{BFY} and in certain cosmological models
\cite{YY}.  

\subsection{Derivation of the Chiellini condition for integrability}

Recently Harko \textit{et al} \cite{Harko1, Harko2} have considered
certain exactly integrable cases of the Li\'{e}nard equation by
appealing to an integrability criterion known as the Chiellini
condition and making use of the first-order Abel equation.
 Multiplying (\ref{b}) by $\exp(-\int g_2(x)
dx)$ and  setting $u=v\exp(-\int g_2(x) dx)$ leads us to the
standard form of the Abel equation of the first kind namely
\cite{CRS}\be\label{c} \frac{du}{dx}=A(x)u^2+B(x)u^3,\ee where
$A(x)=g_1(x)\exp(\int g_2(x) dx)$ and $B(x)=g_0(x) \exp(2\int
g_2(x) dx)$ respectively. An exact solution of (\ref{c}) can be
constructed
 provided the Chiellini condition for integrability
\cite{Chiellini, Kamke} for (\ref{c}),  given by
\be\label{Che1}\frac{d}{dx}\left(\frac{B}{A}\right)=sA(x),\ee  is
satisfied where $s$ is a nonzero constant \cite{Harko1}. When
$g_1(x)\ne 0$, the above condition becomes
\be\label{Che2}\frac{d}{dx}\left(\frac{g_0}{g_1}\right)=sg_1-\frac{g_0g_2}{g_1}.\ee
 In order to derive this condition let us
consider the following  generalized version of the
Li\'enard  equation, \textit{viz}
\be\label{Lien} \ddot{x} + g_n(x)\dot{x}^{n} +
g_0(x) = 0. \ee
Set $\dot{x} = \xi(x)$, so that (\ref{Lien}) becomes
\be\label{Lien1} \xi {\xi}^{\prime} + g_n(x)\xi^n + g_0(x) = 0, \qquad
\xi^{\prime} = \frac{d\xi}{dx}. \ee Suppose $\xi = F(x)G(u(x))$,
where $G$ is a function of $u$. By differentiating $\xi$ with
respect to $x$ and substituting it back into (\ref{Lien1}) we
obtain \be\label{Lien2} {u}^{\prime}=-\frac{\displaystyle
FF^\prime G^2+g_n(x)F^n G^n +g_0(x)}{F^2G\frac{\partial G}{\partial
u}}. \ee In order to separate the variables and integrate equation
(\ref{Lien2}), we observe that the function $F$ should satisfy:
$$
\frac{{F}^{\prime}}{F} = k g_n(x)F^{n-2} = l \frac{g_0(x)}{F^2},
$$
where $k$ and $l$ and constants or in other words
$$kg_n(x)=\frac{F^\prime}{F^{n-1}}, \;\;lg_0(x)=F F^\prime.$$  From these relations we obtain
$$F^{2-n}=(2-n)\int kg_n(x) dx\;\;\;\mbox{and}\;\;\frac{l}{k}\frac{g_0}{g_n}=F^n$$
whence we have
\be\label{Lien2a} \frac{d}{dx}\left(\frac{g_0}{g_n}\right)=\frac{k^2}{l}(2-n)g_n(x)\left((2-n)k\int g_n(x) dx\right)^{2(n-1)/(2-n)}.\ee

Now suppose $G = u$, then
(\ref{Lien2}) reduces to \be\label{Lien3} {u}^{\prime} = -k
g_n(x)F^{n-2}F^{n-2}u - g_n(x)F^{n-2}u^{n-1} - \frac{k}{lu}F^{n-2}
g_n(x) = -g_n(x)F^{n-2} \left(ku + u^{n-1} + \frac{k}{lu}\right). \ee
This being separable it is solvable.\\

Setting $n=1$, (\ref{Lien2a}) reduces to
\be\label{Lien4}\frac{d}{dx}\left(\frac{g_0}{g_1}\right)=\frac{k^2}{l}g_1(x),\ee
while from (1) it follows that when $g_2=0$,
$$\frac{dv}{dx}=g_1v^2+g_0v^3$$ which is to be compared with (\ref{c}). It is now obvious that the criterion stated in (\ref{Che1}) is identical to (\ref{Lien4}) with $s=k^2/l$.


\subsection{Construction of an implicit solution}

As explained in \cite{Harko1} an implicit solution of (\ref{c}) can be accomplished by defining a new variable $w=uB/A$ and using the  Chiellini condition such that (\ref{c}) is transformed to
\be\label{c1}\frac{dw}{dx}=\frac{A^2}{B}w(w^2+w+s).\ee
This leads to a separation of the variables, namely
\be\label{c2}
F(w,s):=\int \frac{dw}{w(w^2+w+s)}=\int\frac{A^2}{B} dx=\frac{1}{s}\int d\ln(B/A)\ee where the  Chiellini
condition has been used once again and finally allows us to express the solution of (\ref{c1}) in the implicit form
\be\label{c3}\Big|\frac{B}{A}\Big|=K^{-1}e^{sF(w,s)}\ee where $K^{-1}$ is an arbitrary constant of integration.
 It follows that
 $$\dot{x}=\frac{1}{v}=\frac{1}{ue^{\int g_2 dx}}=\frac{B}{Ae^{\int g_2 dx}w(x)}=\frac{g_0(x)}{g_1(x)w(x)}$$
 and hence
 \be t-t_0=\int \frac{w(x)g_1(x)}{g_0(x)} dx.\ee
 The form of the right-hand-side of (\ref{c3}) depends on the value of the parameter $s$ and
 \be\label{c5}e^{sF(w, s)}=\left\{\begin{array}{cc}
 \frac{w}{\sqrt{w^2+w+s}}
 \exp\left(-\frac{1}{\sqrt{4s-1}}\arctan\left(\frac{2w+1}{\sqrt{4s-1}}\right)\right)&s>\frac{1}{4}\\
 \frac{w}{w+\frac{1}{2}}\exp\left(\frac{1}{2w+1}\right)&s=\frac{1}{4}\\
 \frac{w}{\sqrt{w^2+w+s}}\Big|1-\frac{1+2w}{\sqrt{1-4s}}\Big|^{-\frac{1}{2\sqrt{1-4s}}}
 \Big|1+\frac{1+2w}{\sqrt{1-4s}}\Big|^{\frac{1}{2\sqrt{1-4s}}}& s<\frac{1}{4}\end{array}\right.\ee

\section{Solution of first-order ODEs via an auxiliary system of
ODEs}
In \cite{SJKB} the behaviour of the dynamical system described by the equation
\be\label{x1}\ddot{x}+\alpha x^{2n+1}\dot{x}+x^{4n+3}=0,\ee was analysed and it was conjectured that there exists
a critical value of the parameter $\alpha_c$ below  which the system admits closed orbits. Extensive numerical computations indicated that the critical value was $\alpha_c=2\sqrt{2(n+1)}$. In view of the method described above one can obtain analytically this critical value by reducing the equation to a first-order Abel equation:
\be\label{x2}\frac{dv}{dx}=\alpha x^{2n+1}v^2+x^{4n+3}v^3.\ee
It is observed that the Cheillini integrability condition is satisfied with the constant $s=2(n+1)/\alpha^2$. From (\ref{c5}) it is seen that the nature of the solution for $w$ changes as $s$ varies from less than $1/4$ to greater than $1/4$. The critical value corresponding to $s=1/4$ implies that the parameter $\alpha_c=2\sqrt{2(n+1)}$. This provides a proof of the validity of the conjecture made in \cite{SJKB}.\\
 A useful method of solving a first-order ordinary
differential equation (FOODE) is by the introduction of an auxiliary
system of first-order ODEs which have a common solution with the
given equation \cite{Harko1, Harko2}. To explain how this is
achieved consider a first-order ODE given by
\be\label{X1}
\frac{dv}{dx}=F(x,v),\ee and introduce an auxiliary system of first-order
ODEs \be\label{X2a}\frac{dv}{dx}=-F_1(x,v)+G(x)f(v),\ee
\be\label{X2b}\frac{dv}{dx}=\frac{1}{2}F_2(x,v)+\frac{1}{2}G(x)f(v),\ee
subject to the constraint $F_1+F_2=F$, where $G(x)$ is a function
to be determined. If a function $G(x)$ exists such that
(\ref{X2a}) and (\ref{X2b}) have a common solution then it is
easy to show that this solution satisfies the equation (\ref{X1}).
The above technique can be adapted to deal with second-order
ODEs which frequently arise in   physical applications.\\

Consider  a second-order ODE of the form (\ref{a}), \textit{viz}
\be\label{S1}\frac{d^2x}{dt^2}+g_1(x)\frac{dx}{dt}+g_2(x)\left(\frac{dx}{dt}\right)^2+g_0(x)=0.\ee
Typically if $g_2=0$ then we have an equation of the Li\'{e}nard
type,  and if $g_1=0$ we obtain an equation with a quadratic dependance on the velocity which, from a Newtonian point of view, may be interpreted as arising from the dependance of the mass of a particle on its position coordinate.
Both types of equations having either a linear or a quadratic dependance on the velocity have been extensively studied
 \cite{CG, GC, RC, Sabatini}.  The transformation
$dx/dt=1/v(x)$ causes (\ref{S1}) to become
\be\label{S2}\frac{dv}{dx}=g_0 v^3+g_1v^2+g_2 v:=F(x, v).\ee
Demanding $F_1=F(x,v), F_2=0$ and $f(v)=v^3$ the analogs of
(\ref{X2a}) and (\ref{X2b}) then have the following forms, in terms
of the transformed variables, namely:
\begin{align}
\label{X3a}\frac{dv}{dx} &=(G(x)-g_0)v^3-g_1v^2-g_2v,\\
\label{X3b}\frac{dv}{dx}&=\frac{1}{2}G(x)v^3.\end{align} The use of
the Chiellini condition for (\ref{X3a}) allows us to express
$G(x)$ as \be\label{X4}G(x)=g_0+g_1\exp(\int
g_2 dx)\left[\Gamma +s^\prime\int g_1 \exp\left(-\int g_2 dx\right)
dx\right],\ee where $\Gamma$ is a constant of integration with the
constant $s^\prime$ appearing as a result of the use of the Chiellini
condition.  Notice that owing to the  convenient choices made for
the functions $F_1$ and $F_2$, (\ref{X3b}) is separable and its
solution is  given by \be\label{X5}\frac{1}{v}=\frac{dx}{dt}=\pm
\sqrt{B-\int G(x)
dy},\ee with $B$ being a constant of integration. \\
Now the existence of a common solution means that
$$\frac{dv}{dx}=\frac{1}{2}G(x)v^3=(G(x)-g_0)v^3-g_1 v^2-g_2 v,$$
which implies upon using (\ref{X5})
\be\label{X6}\frac{G(x)-2g_0}{g_1}=\pm2\sqrt{B-\int G(x)
dx}+\frac{2g_2}{g_1}\left(B-\int G(x) dx\right).\ee Eqn (\ref{X6})
may be used to determine the values of the parameters $s^\prime, \Gamma$
and $B$ after substituting the value of $G(x)$ from (\ref{X4}).
Knowledge of  $G(x)$ then allows us to obtain the common solution  from (\ref{X5}) in
the  form \be\label{X7}\pm t-t_0=\int
\frac{dx}{\sqrt{B-\int G(x) dx}},\ee with $t_0$ being a parameter
which defines the families of solutions. The procedure is illustrated
 below.

 \noindent {\textbf{Example 3.1:}} $\ddot{x}+\alpha x^{2n+1}\dot{x}+x^{4n+3}+w_0^2 x^{2n+1}=0$\\
 Under the transformation $\dot{x}=1/v$ this equation becomes
 $$\frac{dv}{dx}=(\alpha x^{2n+1}) v^2 +(x^{4n+3} +w_0^2 x^{2n+1}) v^3:=F(x,v)$$
 Choose the auxiliary system of FOODEs to be the following:
 $$\frac{dv}{dx}=-F(x,v) +G(x) v^3=(G(x)-x^{4n+3}-w_0^2 x^{2n+1})v^3-\alpha x^{2n+1} v^2$$
 $$\frac{dv}{dx}=\frac{1}{2} G(x) v^3$$
 Applying the Cheillini integrability condition to the first of these equations we have
 $$\frac{d}{dx}\left(\frac{G(x)-x^{4n+3}-w_0^2 x^{2n+1}}{-\alpha x^{2n+1}}\right)=s^\prime (-\alpha x^{2n+1})$$
 We solve this for $G(x)$ to get
 \be\label{A1}G(x)=x^{4n+3}\left(\frac{s^\prime \alpha^2}{2(n+1)}+1\right)+(\alpha \Gamma +w_0^2)x^{2n+1}\ee
 Upon solving the second auxiliary FOODE, which is separable, we obtain
 \be\label{A1.1}\frac{1}{v}=\pm \sqrt{B-\int G(x) dx}\ee
 where $B$ and $\Gamma$ are arbitrary constants of integration. If a common solution exists for the two auxiliary FOODEs then we must have
 $$\frac{1}{2}G(x) v^3=(G(x)-x^{4n+3}-w_0^2 x^{2n+1})v^3+(-\alpha x^{2n+1})v^2$$
 which leads
\be\label{A2}G(x)=2x^{2n+1}\left[(x^{2(n+1)}+w_0^2)\pm \alpha\sqrt{B-\int G(x) dx}\right]\ee
 Equating (\ref{A1}) and (\ref{A2}) we have upon equating coefficients of different powers of $x$ (with
 $\xi=\alpha^2 s^\prime/2(n+1)-1$),
 \be\label{B1}\xi^2=-\frac{\alpha^2}{n+1}(\xi+2)\ee
 \be\label{B2}\xi(\alpha\Gamma-w_0^2)=-\frac{\alpha^2}{n+1}(\alpha\Gamma+w_0^2),\;\;\;
 (\alpha\Gamma-w_0^2)^2=4\alpha^2 B\ee
 We can solve for the constants of integration and $\xi$ to obtain
 \be\label{B3}\xi=\frac{1}{2}\left[-\left(\frac{\alpha^2}{n+1}\right)\pm \sqrt{\left(\frac{\alpha^2}{n+1}\right)^2-
 8\left(\frac{\alpha^2}{n+1}\right)}\right]\ee
 \be\label{B4}\Gamma=\frac{w_0^2}{\alpha}\left[\frac{(n+1)\xi-\alpha^2}{(n+1)\xi +\alpha^2}\right],\;\;\;
 B=\frac{\alpha^2w_0^4}{[(n+1)\xi +\alpha^2]^2}\ee
 Knowing the constants of integration,  the solution may be reduced to  quadrature using (\ref{A1}) and (\ref{A1.1}), i.e.,
 \be\label{A.3}\pm t-t_0=\int \frac{dx}{\sqrt{B-\frac{\xi+2}{4(n+1)}x^{4(n+1)}-\frac{w_0^2\xi}{[(n+1)\xi+\alpha^2]}x^{2(n+1)}}}.\ee
 \textbf{Case A:} $\alpha=2n+3$ and $w_0\ne 0$\\
 For this choice of the parameter $\alpha$ we find from (\ref{B3}) since $s^\prime=2(n+1)(\xi+1)/\alpha^2$
 $$(s_+^\prime, s_-^\prime)=\left(-\frac{2(n+2)}{(2n+3)^2}, -\frac{2(n+1)(4n+5)}{(2n+3)^2}\right)$$
 $$(B_+, B_-)=\left(\frac{w_0^4}{4(n+1)^2}, w_0^4\right)$$
 $$(\Gamma_+, \Gamma_-)=\left(-\frac{(n+2)}{(n+1)(2n+3)}w_0^2, -\frac{(4n+5)}{(2n+3)}w_0^2\right)$$
 These values lead to the following expressions for the unknown function $G(x)$, viz,
 $$G_+(x)=-\frac{1}{n+1}\left[x^{4n+3}+w_0^2x^{2n+1}\right]$$
 $$G_-(x)=-4(n+1)\left[x^{4n+3}+w_0^2x^{2n+1}\right]$$
 which in turn yield the solutions
 $$\pm t-t_0^+=2(n+1)\int \frac{dx}{x^{2(n+1)}+w_0^2},$$
 $$\pm t-t_0^-=\int \frac{dx}{x^{2(n+1)}+w_0^2}$$
 respectively.

\begin{figure}[h]

\centering
\includegraphics[width=.7\linewidth]{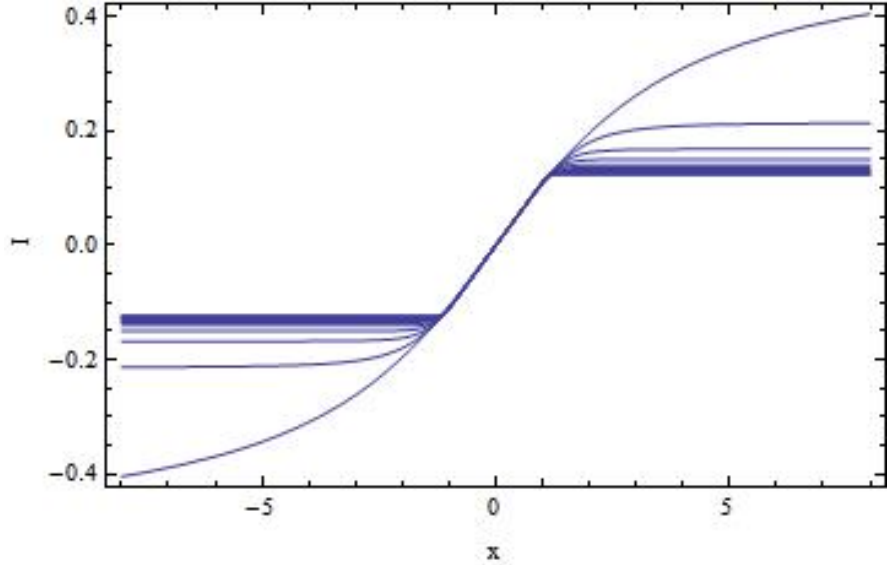}
\caption{Graph for $n=0 -10$, for $n=0$ we obtain arctan and for large $n$ curves are dense}.
\end{figure}

 It is evident from these solutions that they are equivalent up to a scaling. Indeed setting $w_0=1$ one may explicitly express the solution in terms of the hypergeometric function  $_2F_1(a,b;c; x)$ because
 $$\int \frac{dx}{x^{2(n+1)}+1}=x  _2F_1(1, \frac{1}{2(n+1)}; 1+\frac{1}{2(n+1)}; -x^{2(n+1)})$$
 From (\ref{B3}) it is evident that $\alpha^2=(2n+3)^2>8(n+1)$. The critical value of $\alpha$ corresponding to the vanishing of the discriminant is $\alpha_c=2\sqrt{2(n+1)}$. Thus when $\alpha>\alpha_c$  and $n=0$ we obtain the case of periodic motion. Incidently this corresponds to isochronous motion, in which the period function is independent of the initial condition. This is easily verified from the corresponding criterion given by Sabatini in \cite{Sabatini}.
 There it is shown that for a Li\'{e}nard equation
 $\ddot{x}+f(x)\dot{x}+g(x)=0$ having an isochronous center at the origin  with $f, g\in C^{1}(J, R), f(0)=g(0)=0, g^\prime(0)>0$ the forcing term $g(x)$ must be of the form
 $$g(x)=g^\prime(0)x+\frac{1}{x^3}\left(\int_0^x sf(s)ds\right)^2$$
It is straightforward to verify that these conditions are satisfied by the equation
$\ddot{x} + (2n + 3)x^{2n+1}\dot{x} + x^{4n+3} + w_0^2x = 0$,
and hence by the equation $\ddot{x} + 3x\dot{x} + x^3 + w_0^2x = 0$, which corresponds to $n=0$.\\

 \noindent
 \textbf{Case B:} $\alpha=2n+3$ and $w_0= 0$\\
 When $w_0=0$  we have from (\ref{B4}) that $\Gamma=B=0$ and the solutions are
 $$x=\left[\frac{2(n+1)}{2n+1}\frac{1}{(t_0^+\mp t)}\right]^{1/(2n+1)}\;\;\;\mbox{and}\;\;\;
 x=\left[\frac{1}{2n+1}\frac{1}{(t_0^-\mp t)}\right]^{1/(2n+1)}$$
 respectively and are singular. \\

\begin{figure}[h]

\centering
\includegraphics[width=.7\linewidth]{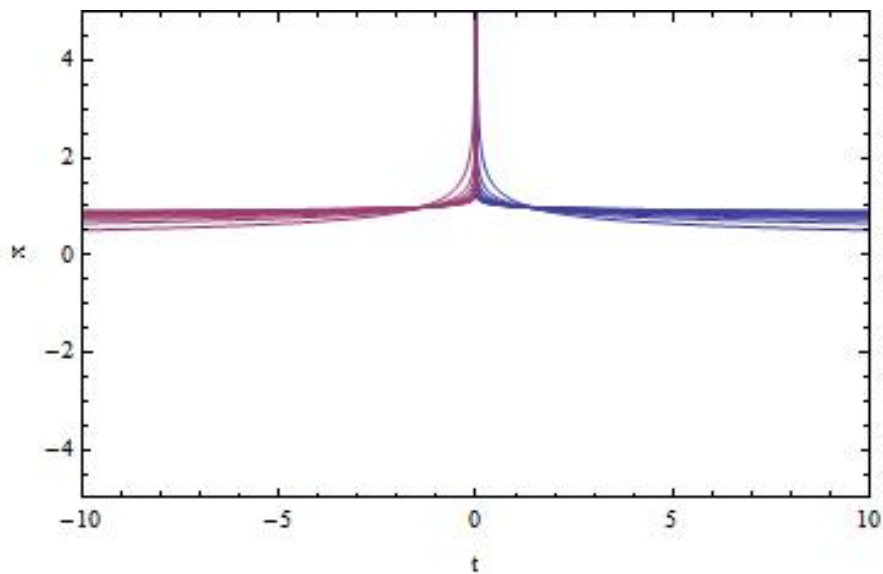}
\caption{Graph for $n=1-10$, where $t_0^-$ and $t_0^+$ are approaching from lhs and rhs of $t=0$}.
\end{figure}

 \section{Generalizations of the Chiellini condition}
 The Chiellini
integrability condition has been used in a number of works (see
\cite{Harko2} and references therein). Its generalization to the
case when higher powers of $u$ appear in the right hand side of
(\ref{c}) has also been studied. In view of it efficacy in deriving solutions of the first-order Abel equation
 we consider below higher-order generalizations of  the Li\'{e}nard equation.

\subsection {Higher-order Li\'enard equation}
 Consider the higher-order  Li\'enard equation
 \be\label{gLien} \ddot{x} +
f(x)\dot{x}^{n+1} + g(x)\dot{x}^{n} = 0. \ee Suppose $\dot{x} =
\xi(x)$, so that (\ref{gLien}) becomes \be\label{gLien1}
{\xi}^{\prime} + f(x)\xi^n + g(x)\xi^{n-1} = 0. \ee Once again we
assume $\xi = F(x)G(u(x))$, where $G$ is a function of $u$.
Following the procedure outlined in Section 4.1 we obtain
\be\label{gLien2} {u}^{\prime} =-\frac{\displaystyle F^\prime
F^{n-3}G^n +f(x)F^{2n-3}G^{2n-1}
+g(x)}{F^{n-2}G^{n-1}\frac{\partial G}{\partial u}}.
 \ee After
separating the variables we have
$$
\frac{{F}^{\prime}}{F} = k f(x)F^{n-1} = l \frac{g(x)}{F^{n-2}},
$$
and this leads to the {\it generalized  Chiellini condition}
\cite{Harko2} \be\label{gLien3} \left(\frac{g}{f}\right)^{\prime}
= \frac{l^{n-1}}{k^{n-2}}\left(\frac{g^n}{f^{n-1}}\right) \equiv
K\left(\frac{g^n}{f^{n-1}}\right), \ee
 which for $n = 0$  reduces to the usual  Chiellini condition stated in (\ref{Lien4}).
Upon introducing the transformation $$ \xi =
\left(\frac{g(x)}{f(x)}\right)\eta(x), $$  (\ref{gLien1}) becomes
\be\label{gLien4} \frac{d\eta(x)}{dx} =
\frac{g^{n-1}(x)}{f^{n-2}(x)} \left(\eta^n + \eta^{n-1} + K\eta
\right), \ee which is clearly separable.

 \section*{Acknowledgement}

The authors wish to thank Professors J. K Bhattacharjee and  A. Mallik for their interest and encouragement. 
One of us (PG) wishes to acknowledge Professor Tudor Ratiu for his gracious hospitality at the
Bernoulli Centre, EPFL during  the fall semester of 2014, where part of this work was done.

\end{document}